\documentclass[a4paper,runningheads]{llncs}

\usepackage[T1]{fontenc}
\usepackage{lmodern}

\usepackage{sosy-paper}
\usepackage{changepage}
\usepackage{syntax}
\usepackage{balance}
\usepackage{longtable}
\usepackage{rotating}
\usepackage{colortbl}
\usepackage{textcomp}
\usepackage{listings}
\usepackage[paperwidth=155mm,
            paperheight=235mm,
            left=16mm,
            right=16mm,
            top=17mm,
            bottom=25mm
           ]{geometry}

\title{Correctness Witnesses with Function Contracts}

\author{
    Matthias~Heizmann\texorpdfstring{\orcidID{0000-0003-4252-3558}\inst{1}}{},
    Dominik~Klumpp\texorpdfstring{\orcidID{0000-0003-4885-0728}\inst{2}}{}, \\
    Marian~Lingsch-Rosenfeld\texorpdfstring{\orcidID{0000-0002-8172-3184}\inst{3}}{},
    Frank~Schüssele\texorpdfstring{\orcidID{0000-0002-5656-306X}\inst{2}}{}
}

\authorrunning{M. Heizmann, D. Klumpp, M. Lingsch-Rosenfeld, F. Schüssele}

\institute{
    University of Stuttgart, Stuttgart, Germany \\
    \email{matthias.heizmann@iste.uni-stuttgart.de}
\and 
    University of Freiburg, Freiburg, Germany \\
    \email{\{klumpp,schuessf\}@informatik.uni-freiburg.de}
\and 
    LMU Munich, Munich, Germany \\
    \email{marian.lingsch-rosenfeld@sosy.ifi.lmu.de}
}

\DisableLigatures{family=tt*}

\newcommand{\ttbox}[1]{\colorbox{black!8}{\footnotesize\texttt{#1}\tiny\strut}}
\newlength\multilen
\newcommand{\multi}[2][\multilen]{\parbox[c]{#1}{\baselineskip=0pt \vspace{2.5pt}\strut #2 \vspace{2.5pt}}}
\newcommand{\opt}{${}^*$}

\makeatletter
\def\myrulefill{\leavevmode\leaders\hrule height .7ex width 1ex depth -0.6ex\hfill\kern\z@}
\makeatother
\newcommand{\separator}[1]{\multicolumn{3}{c}{\myrulefill\myrulefill\quad\!\!\!#1\vphantom{\large $\mid$}\!\!\!\quad\myrulefill\myrulefill}}

\newcommand{\examplecodesizett}{\fontsize{7}{8.5}\ttfamily\selectfont}

\lstdefinestyle{cacsl}{
  commentstyle=\color{green}\itshape,
  morekeywords={[2]assert,nondet,nondet_short,nondet_int,nondet_unsigned,product,main,increment},
  keywordstyle=[2]\color{blue}
}

\newcommand\YAMLcolonstyle{\color{magenta}\mdseries}
\newcommand\YAMLkeystyle{\color{blue}\examplecodesizett}
\newcommand\YAMLvaluestyle{\color{black}\examplecodesizett}
\makeatletter

\newcommand\language@yaml{yaml}

\expandafter\expandafter\expandafter\lstdefinelanguage
\expandafter{\language@yaml}
{
keywords={true,false,null,y,n,{...}},
keywordstyle=\color{darkgray}\bfseries,
basicstyle=\YAMLkeystyle,                                 %
sensitive=false,
comment=[l]{\#},
morecomment=[s]{/*}{*/},
commentstyle=\color{gray}\ttfamily,
stringstyle=\YAMLvaluestyle\ttfamily\color{green},
showstringspaces=false,
moredelim=[l][\color{orange}]{\&},
moredelim=[l][\color{magenta}]{*},
moredelim=**[il][\YAMLcolonstyle{:}\YAMLvaluestyle]{:},   %
morestring=[b]',
morestring=[b]",
literate = {...}{{{\hspace{1em}\color{gray}...}}}1
}

\lst@AddToHook{EveryLine}{\ifx\lst@language\language@yaml\YAMLkeystyle\fi}
\makeatother

\definetool{\vercors}{VerCors}
\definetool{\verifast}{VeriFast}
\definetool{\dafny}{Dafny}

\newcommand{\resultACSL}{\texttt{\textbackslash result}\xspace}
\newcommand{\old}{\texttt{\textbackslash old}\xspace}
\newcommand{\at}{\texttt{\textbackslash at}\xspace}
\newcommand{\atXPre}[1][x]{\mbox{\texttt{\textbackslash at(#1, Pre)}}\xspace}
\newcommand{\atPre}{\mbox{\texttt{\textbackslash at(\_, Pre)}}\xspace}
\newcommand{\oldX}[1][x]{\texttt{\textbackslash old(#1)}\xspace}
\newcommand{\cExpr}{\texttt{c\_expression}\xspace}
\newcommand{\acslExpr}{\mbox{\texttt{acsl\_expression}}\xspace}
\newcommand{\locInv}{\texttt{location\_invariant}\xspace}
\newcommand{\loopInv}{\texttt{loop\_invariant}\xspace}
\newcommand{\funContract}{\mbox{\texttt{function\_contract}}\xspace}

\newcommand{\hurl}[1]{{\scriptsize\UrlFont\href{https://#1}{#1}}} %
\newcommand{\fhurl}[1]{\footnote{\hurl{#1}}}

\begin{document}

\maketitle

\begin{abstract}
Software verification witnesses are a common exchange format
for software verification tools. They were developed
to provide arguments supporting the verification result,
allowing other tools to reproduce the
verification results.
Correctness witnesses in the current format (version~2.0)
allow only for the encoding
of loop and location invariants using C expressions.
This limits the correctness arguments that verifiers can express in the witness format.
One particular limitation is the inability to express
function contracts, which consist of a pre-condition and a
post-condition for a function.
We propose an extension to the
existing witness format 2.0 to allow for the specification of
function contracts.
Our extension includes support for several features inspired by ACSL
(\resultACSL, \old, \at).
This allows for the export of more information from
tools and for the exchange of information
with tools that require function contracts.


\end{abstract}
\section{Introduction}
\begin{figure}[t]
  \begin{minipage}[t]{0.55\textwidth}
\begin{lstlisting}[language=C,style=cacsl]
/*@ 
  requires b >= 0;
  ensures \result == a * b;
@*/ 
int product(short a, short b) {
  if (b == 0) return 0;
  return a + product(a, b-1);
}

int main() {
  short x = nondet_short();
  short y = nondet_short();
  if (y < 0) return 0;

  int r = 0;
  /*@ 
    loop invariant
      r == x * i &&
      i <= y && y >= 0;
  @*/
  for (int i=0; i<y; i++) {
    r += x;
  }
  assert(r == product(x, y));
}
\end{lstlisting}
  \end{minipage}
  \begin{minipage}[t]{0.45\textwidth}
\begin{lstlisting}[language=C,style=cacsl]
int g;

/*@
  requires g > 0;
  ensures g < \old(g);
@*/
void div() {
  g = g / 2;

  /*@ 
    loop invariant 
      g < \at(g, Pre) &&
      g >= 0;
  @*/
  while (nondet_int()) {
    g = g / 2;
  }
}

int main() {
  int x = nondet_int();
  g = x;
  if (g > 0) {
    div();
    assert(g < x);
  }
}
\end{lstlisting}
  \end{minipage}
  \caption{C programs annotated with (relational) function contracts and (relational) invariants in ACSL.}
  \label{fig:example-acsl}
\end{figure}

Software verification witnesses~\cite{Witnesses,VerificationWitnesses-2.0}
were developed
in the context of the Software Verification Competition
(SV-COMP)~\cite{SVCOMP15}.
Witnesses contain arguments supporting the verification result,
allowing other tools to independently reproduce the verification results.
This is called \emph{validation} of the verification results.
The information provided in a witness is believed to be what  
allows validators to usually reproduce
the results faster than verification from scratch
and helps to catch bugs in verification tools~\cite{Validators}.

Nowadays, software verification witnesses are a widely adopted
common exchange format for software verification tools,
providing a machine-readable interface to interact with
them.
They do not allow only for the validation of results, but also
for the exchange of information between different tools
with the goal of cooperatively solving a
problem~\cite{CooperativeVerification,CCEGAR,InterfaceTheoryProgramVerification,CEGAR-PT}.
In particular, there has been work on the exchange of information with deductive
verifiers~\cite{AutoActive}.

Correctness witnesses in version 2.0~\cite{VerificationWitnesses-2.0}
allow only for the encoding of loop and location invariants
using C expressions.
This significantly limits the arguments verifiers can express
for the correctness of a program. One particular
limitation is the inability to express function contracts,
which consist of a pre- and post-condition of a function.
Function contracts are a widely established formalism
for modular verification of software
systems~\cite{DesignByContract, DeductivePenAndPaper}.
In the same manner as functions allow for the modularization
of code, function contracts allow for the modularization
of a proof. Due to this, they are one of the most used methods to specify the behavior of interprocedural programs,
e.g.,\ in \emph{design by contract}~\cite{DesignByContract} or
by deductive verifiers~\cite{Dafny,FramaC,VeriFast,VerCorsTool}.

As an example, consider the function \texttt{product} in~\cref{fig:example-acsl} (left)
which has been annotated with a function contract in ACSL notation~\cite{ACSL}.
The \texttt{requires} clause expresses that the function must only be called with a non-negative second argument \texttt{b}.
The \texttt{ensures} clause states that the return value (denoted by \resultACSL) is the product of the arguments \texttt{a} and \texttt{b}.
This information is crucial to prove the correctness of the program,
i.e., that the \texttt{assert} in line~24 can never fail.
Together with the loop invariant in line~17,
this annotation allows a \emph{deductive verifier} such as \framac~\cite{FramaC},\vercors~\cite{VerCorsTool}, \dafny~\cite{Dafny} or \verifast~\cite{VeriFast} to prove the program's correctness.
As another example, consider the function \texttt{div} in~\cref{fig:example-acsl} (right). It requires a function contract
which relates the value of the global variable \texttt{g} before and after the function call (denoted by \oldX[g]) to express the correctness of the program, i.e.,
that the assertion in line~25 always holds. Additionally,
to show that the program is correct,
the loop invariant in line~11 requires a relation between
the variable in the loop \texttt{g}
to its value before the function call (denoted by \atXPre[g]).

We propose an extension to the
existing witness format 2.0 to allow for the encoding of
function contracts.
In order to specify meaningful function contracts,
we additionally permit function contracts and invariants to use the expression format \acslExpr (rather than plain C expressions).
This ACSL-inspired format allows for the usage of the keywords \old and \atPre
for function contracts and invariants respectively to refer to the values of global variables or parameters
at the beginning of the current function call,
and \resultACSL to refer to the return value of a function.
This extension increases the
expressiveness of the witness format and allows for the exchange
of function contracts.
In particular, the proposed extension fulfills our
design goals of \emph{compatibility} with existing tools
and \emph{increased expressiveness}.
We present the format in~\cref{sec:syntax} with its
semantics in~\cref{sec:semantics}. Finally, we discuss
some implications of this extension and
how it fulfills our design goals in~\cref{sec:discussion}.
\section{Syntax}
\label{sec:syntax}

To express function contracts in software verification
witnesses, we extend the format
version 2.0~\cite{VerificationWitnesses-2.0} by
adding a new entry type \funContract to the
allowed entries inside an \texttt{invariant_set}.
\Cref{tab:function-contract} shows the contents of the new entry
for function contracts. As with \locInv and \loopInv,
it is inside a mapping with the key \texttt{invariant} to
remain consistent with the existing format.
Furthermore, the existing entry types \locInv and \loopInv
in the \texttt{invariant_set} are extended by allowing the new format \acslExpr to allow for the use of the ACSL construct \atPre.

\begin{table}[t]
  \renewcommand{\arraystretch}{1.05}
  \renewcommand{\opt}{\textsuperscript{?}}
  \setlength{\tabcolsep}{3.5pt}
  \setlength{\multilen}{6.5cm}
  \caption{Structure of an \ttbox{invariant} entry for function contracts;
    optional items are marked with \opt}
  \centering
  \begin{tabular}{lll}
    \toprule
    Key & Value & Description \\ \midrule
    \ttbox{invariant} & mapping & \multi{the entry containing all information pertaining to a function contract; see below}\\
    \separator{content of \ttbox{invariant}} \\
    \ttbox{type} & \ttbox{function_contract} & \multi{the type of the newly introduced entry} \\
    \ttbox{location} & mapping & the location of the function contract; see below\\
    \ttbox{format} & \ttbox{c_expression} & \multi{all elements of the function contract are C expressions} \\
                   & \ttbox{acsl_expression} & \multi{all elements of the function contract are ACSL expressions. This is always the case when \\ keywords like \old and \resultACSL are used.} \\
    \ttbox{requires}\opt & scalar & \multi{expresses the pre-condition of the function, \\ defaults to \texttt{true} when not present} \\
    \ttbox{ensures}\opt & scalar & \multi{expresses the post-condition of the function, defaults to \texttt{true} when not present} \\
    \separator{content of \ttbox{location}} \\
    \ttbox{file_name} & scalar & the file of the location\\
    \ttbox{line} & scalar & the line number of the location\\
    \ttbox{column}\opt & scalar & the column of the location\\
    \ttbox{function}\opt & scalar & the name of the function \\
    \bottomrule
  \end{tabular}
  \label{tab:function-contract}
\end{table}

A \funContract contains the
\texttt{requires} and \texttt{ensures}
clauses which express the pre-condition and post-condition
of the function.
Both clauses are optional with a default value of \texttt{1} (the C expression representing $\mathit{true}$).
The function contract's \texttt{location} points to the first character
of the function definition. For example, for the function definition \texttt{int main(...) \{...\}} the location should point to the \texttt{i} of \texttt{int}.
If the column
is missing, the location points to the first allowed position
in the line i.e. to the first character of the first
function definition on that line.
Multiple function contracts are allowed for the same function
and all need to be valid for the witness to be valid.

For a function contract with the format \cExpr,
the clauses \texttt{requires} and \texttt{ensures} must be syntactically correct, side effect-free C expressions.
The expressions may only use global variables and parameters of the function.
If the format \acslExpr is used instead,
\texttt{ensures} may additionally contain the expression \oldX,
where \texttt{x} is a global variable or parameter,
and the keyword \resultACSL.
The expression \oldX refers to the value of \texttt{x} at the beginning of the function call. In contrast to ACSL, we only allow variables
as arguments to \old, not arbitrary expressions.
The keyword \resultACSL refers to the function's return value.
As such, if the function's return type is \texttt{void}, the keyword \resultACSL must not be used.

The proposed expression format \acslExpr can also be used in the existing entries \locInv and \loopInv.
Invariants with this expression format may contain any expression that is valid in the expression format \cExpr,
as well as expressions of the form \atXPre, where \texttt{x} is a global variable or a parameter of the function. The expression
\atXPre refers to the value of \texttt{x} at the beginning of the execution of the current function call.
Note that, while ACSL allows \at expressions to be used with labels in the program text as well as \emph{logic labels} other than \texttt{Pre}, we allow only \atPre in our witness format. 

These changes have been proposed in the SV-Witnesses
repository\fhurl{gitlab.com/sosy-lab/benchmarking/sv-witnesses/-/merge\_requests/87}, by adding
a new schema for correctness witnesses and implementing
their linting.
Additionally, further examples can also be found there.

\Cref{fig:example-witness} shows the left program from \cref{fig:example-acsl} again.
Here, the ACSL annotations have been omitted.
Instead, the function contract and the loop invariant are encoded in the correctness witness on the right, using the proposed format.

\begin{figure}[t]
  \begin{minipage}[t]{0.49\textwidth}
\begin{lstlisting}[language=C,style=cacsl]
int product(short a, short b) {
  if (b == 0) return 0;
  return a + product(a, b-1);
}

int main() {
  short x = nondet_short();
  short y = nondet_short();
  if (y < 0) return 0;

  int res = 0;
  for (int i=0; i<y; i++) {
    res += x;
  }
  assert(res == product(x, y));
}
\end{lstlisting}
  \end{minipage}
  \begin{minipage}[t]{0.5\textwidth}
\begin{lstlisting}[language=yaml,numbers=none,breaklines=true]
- entry_type: invariant_set
  metadata: ...
  content:
  - invariant:
      type: loop_invariant
      location:
        file_name: ...
        line: 12
      value: 'res == x * i && i <= y && y >= 0'
      format: c_expression
  - invariant:
      type: function_contract
      location:
        file_name: ...
        line: 1
      requires: 'b >= 0'
      ensures: '\result == a * b'
      format: acsl_expression
\end{lstlisting}
  \end{minipage}
  \caption{An example program (left) and a excerpt of a correctness witness with function contracts (right)}
  \label{fig:example-witness}
\end{figure}
\section{Semantics}
\label{sec:semantics}

A correctness witness in our extended format is valid if and only if each entry of type \funContract, \loopInv and \locInv in the witness is valid and
no program execution results in a violation of the specification.

An entry of type \funContract is valid if and only if its \texttt{requires} clause and its \texttt{ensures} clause are valid. This is trivially the case if the clauses are not present, as they default to \texttt{true}.
A clause \texttt{requires expr} in a contract for a function \texttt{f} is valid
if and only if \texttt{expr} evaluates to a non-zero value before the first statement of the body of the definition of \texttt{f} for every function call of \texttt{f} in in every program execution starting in an initial state, i.e., starting in \texttt{main}.
A clause \texttt{ensures expr} in a contract for the function \texttt{f} is valid
if and only if \texttt{expr} evaluates to a non-zero value after every \texttt{return} statement (i.e., after the returned expression is evaluated) of \texttt{f} before returning to the caller for every function call of \texttt{f} in every program execution that started in an initial state, i.e., starting in \texttt{main}.

For function contracts with format \cExpr and for \texttt{requires} clauses regardless of the format, the evaluation of expressions follows the C~standard as usual.
If the format of a function contract for the function \texttt{f} is \acslExpr,
then the expression in the \texttt{ensures} clause is evaluated the following way:
\begin{itemize}
  \item \oldX evaluates to the value of the global variable or parameter \texttt{x} before the first statement of the function body was executed.
  \item A parameter \texttt{x} always evaluates to the value before the call, i.e., \texttt{x} evaluates to the same as \oldX.
  \item The expression \resultACSL evaluates to the return value of \texttt{f}.
  \item All other expressions are evaluated in the same way as for the expression format \cExpr.
\end{itemize}

The evaluation of a parameter \texttt{x} in an \texttt{ensures} clause is similar to ACSL.
Function contracts describe the externally-visible behavior of the function.
If a parameter is assigned, the new value is not visible to the caller.
Note that, for global variables \texttt{g}, the meaning of \texttt{g} and \oldX[g] in an \texttt{ensures} clause differs.

The validity of the entries \loopInv and \locInv remains largely as in version \texttt{2.0},
i.e., in a program execution that started in an initial state (in \texttt{main}),
they have to hold at every point before the loop condition is evaluated respectively every time  before the execution passes the corresponding location.
The only difference is that, for invariants with expression format \acslExpr,
any sub-expression of the form \atXPre evaluates to the value of the global variable or parameter \texttt{x}
in the pre-state of the current function call, i.e., before the first statement of the function body was executed, similar to function contracts with \oldX.

\section{Discussion}
\label{sec:discussion}

\inlineheadingbf{Compatibility}
The proposed extension of the correctness witness format
is \emph{backwards compatible} for verifiers
and \emph{forwards compatible} for validators.
Every witness in version 2.0 is also a valid witness
in the proposed format.
In particular, it is not required to include a function contract for any function in the witness.
As the \texttt{requires} and \texttt{ensures} clauses default to \texttt{1} when not specified,
they are trivially satisfied by any execution.
As a particular case, the empty witness (which does not contain any \loopInv, \locInv or \funContract entries)
is still a valid witness.
This means that any verifier which exports witnesses
in version 2.0 can also export them in the proposed format.
Conversely, every validator that can validate
correctness witnesses with function contracts
can also validate correctness witnesses in version 2.0.

Adding support for function contracts with the \cExpr format
should be relatively straightforward for most validators.
A \texttt{requires} clause can be handled like a \locInv at the beginning of the function body.
For an \texttt{ensures} clause the situation is similar,
except that the corresponding location -- after evaluation of the returned expression, including potential side-effects, for any \texttt{return} statement or at the end of a function body without \texttt{return} --
does not necessarily correspond to a particular line and column for a statement in the source file.

In order to support function contracts and invariants with expression format \acslExpr,
validators could instrument the program code with local ghost variables that track the value of \oldX, \atXPre and \resultACSL.

\inlineheadingbf{Validation with Deductive Verifiers}
Our extension of the witness format opens the door to using deductive verifiers for witness validation.
Since such tools typically require function contracts for every function in the program,
their usage is severely limited with the current format (2.0).
Witnesses with function contracts allow us to encode full correctness certificates for many programs, in the form of inductive loop invariants and modular function contracts,
which can then be successfully validated by a deductive verifier.

Note however, that the definition of witness validity does not require the given function contracts and invariants to allow for modular verification i.e. they do not necessarily encode a modular proof.
In particular, the pre-condition and post-condition
of a function only need to hold on every program execution
starting in an initial state (in \texttt{main}).
This means that a validator based on a deductive verifier may validate a witness,
but it cannot determine that a witness is invalid.
In case the deductive verifier fails to establish correctness of the program using the witness,
the validator can thus not reject the witness but must output \texttt{unknown}.

\begin{figure}[t]
  \begin{minipage}[t]{0.43\textwidth}
\begin{lstlisting}[language=C,style=cacsl]
int g;

void div() {
  g = g / 2;
  while (nondet_int()) {
    g = g / 2;
  }
}

int main() {
  int x = nondet_int();
  g = x;
  if (g > 0) {
    div();
    assert(g < x);
  }
}
\end{lstlisting}
  \end{minipage}
  \begin{minipage}[t]{0.5\textwidth}
\begin{lstlisting}[language=yaml,numbers=none]
- entry_type: invariant_set
  metadata: ...
  content:
  - invariant:
      type: function_contract
      location:
        file_name: ...
        line: 3
      requires: 'g > 0'
      ensures: 'g < \old(g)'
      format: acsl_expression
  - invariant:
      type: loop_invariant
      location:
        file_name: ...
        line: 5
      value: 'g < \at(g, Pre) && g >= 0'
      format: acsl_expression
\end{lstlisting}
  \end{minipage}
  \vspace{-1.5mm}
  \caption{C program (left) and corresponding witness (right), where \old is necessary to express the correctness of the program.
  The witness encodes the same information
    expressed in \cref{fig:example-acsl} as ACSL annotations.
  }
    \label{fig:needs-old}
    \vspace{-4mm}
\end{figure}

\inlineheadingbf{Increased Expressiveness}
As we have seen (for instance on the examples in~\cref{fig:example-witness}),
function contracts enable us to express correctness arguments that cannot be expressed using the existing witness format.
In particular, the format version 2.0 does not always allow us to specify the \texttt{ensures} clause through one \locInv,
as multiple may be required. For example,
when a function has multiple return points
each one has to be marked with a \locInv to
mimic an \texttt{ensures} clause.
Furthermore, the format only permits expressions over variables in scope at the particular location,
so there is no way to refer to the function return value.
On the other hand, grouping the pre- and post-condition for a function in a contract clearly describes the behavior of the function.
Using the ACSL keywords \resultACSL and \old,
the contract specifies a \emph{relation} between the pre-state and post-state of every function call,
which is not possible with the expression format \cExpr.

\Cref{fig:needs-old} shows another example program with a corresponding witness with function contracts.
For this program, the \old keyword is needed to relate
the value of the global variable \texttt{g} before and after
the function call to express the correctness argument for the program.
Such a relation is needed to conclude that the assertion \texttt{g < x} in line~15 holds,
because the contract cannot refer to the caller's local variable \texttt{x} as it is not in scope.
Further, to prove the relational claim that the function \texttt{div} ensures that the final value of \texttt{g} is smaller than the initial value,
it is also necessary to use \texttt{\textbackslash at(g, Pre)}\xspace in the loop invariant.

\inlineheadingbf{Completeness Remains Challenging}
There still exist correctness arguments for programs which cannot be expressed in the proposed format.
Most notably, invariants that involve the heap and pointers are a major hurdle.
In order to express such invariants,
the expression format needs to be extended further to be able to express, e.g., validity of pointers and the content of arrays.
As our extension is based on ACSL, a mature format for C correctness proofs that has many such features (e.g., a \texttt{\textbackslash valid} keyword and quantifiers),
it paves the way for such further extensions.

Another notable limitation is the lack of data
abstraction, which is crucial for modular verification of programs using complex
data structures. Data abstraction allows us to
create a mathematical model of the data structure,
simplifying the correctness argument. For example,
it may be desired in C to show that a linked list
behaves like a mathematical list, to simplify the
proof.
Existing work~\cite{ContractLib} has already
analyzed how to transfer data abstractions between
implementations using SMT-Lib. A future witness format
which includes data abstractions could be based on this work.

\begin{credits}
    \inlineheadingbf{Acknowledgements} We thank Dirk Beyer, Gidon Ernst, 
    Andreas Podelski and Jan Strejček for their valuable feedback
    and insightful discussions.
    
    \inlineheadingbf{Funding Statement}
    This project was funded in part by the Deutsche Forschungsgemeinschaft (DFG)
    -- \href{http://gepris.dfg.de/gepris/projekt/378803395}{378803395} (ConVeY)
    and \href{https://gepris.dfg.de/gepris/projekt/503812980}{503812980} (IDCC).
\end{credits}

\bibliography{bib/sw,bib/dbeyer,bib/artifacts,bib/svcomp,bib/svcomp-artifacts,bib/testcomp,bib/testcomp-artifacts,bib/websites,bib/new}

\end{document}